\documentclass[pra,amsmath,showpacs,preprint]{revtex4}

\newcommand{\beq}{\begin{equation}}
\newcommand{\eeq}{\end{equation}}
\newcommand{\beqa}{\begin{eqnarray}}
\newcommand{\eeqa}{\end{eqnarray}}
\newcommand{\ket} [1] {\vert #1 \rangle}
\newcommand{\bra} [1] {\langle #1 \vert}
\newcommand{\braket}[2]{\langle #1 | #2 \rangle}
\newcommand{\proj}[1]{\ket{#1}\bra{#1}}

\newcommand{\ketbra}[2]{\ket{#1}\bra{#2}}

\usepackage{ae}
\usepackage[T1]{fontenc}
\usepackage{amsmath}
\usepackage{graphicx}
\usepackage{bm}

\begin{document}
\title{About Covariant Quartit Cloning Machines}

\author{Thomas Durt}
\altaffiliation[Also at: ]{Department of Theoretical Physics (TENA)), Vrije
Universiteit Brussel}
\email{thomdurt@vub.ac.be}
\homepage{http://www.tona.vub.ac.be}
\affiliation{Department of Applied Physics and Photonics (TW-TONA), Vrije
Universiteit Brussel, Pleinlaan 2, B-1050 Brussel, Belgium}

\author{Bob Nagler}
\affiliation{Department of Applied Physics and Photonics (TW-TONA), Vrije
Universiteit Brussel, Pleinlaan 2, B-1050 Brussel, Belgium}

\date{\today}

\begin{abstract} The study of quantum cryptography and quantum entanglement
has traditionally been based on two-level quantum systems (qubits) and more
recently on
three-level systems (qutrits). We investigate several
classes of
state-dependent quantum cloners for four-level systems (quartits). These
results apply to
symmetric as well as asymmetric cloners, so that the balance between the
fidelity of the two
clones can also be analyzed. We extend Cerf's formalism for
cloning states in order to derive cloning machines that remain
invariant under certain unitary transformations. Our results show that
a different cloner has to be used for two mutually unbiased bases
which are related by a double Hadamard
transformation, than for two
mutually unbiased bases that are related by a Fourier transformation. This
different cloner is obtained thanks to a
redefinition of Bell states that respects the intrinsic symmetries of the
Hadamard transformation.
\end{abstract}


\maketitle

\section{Introduction}
Quantum cryptography aims at distributing a random key in such a way that the
presence of an eavesdropper (traditionally called Eve) who monitors the
quantum communication
is revealed by the errors she introduces in the transmission (for a review, see
e.g.~\cite{Gisin02}). To realize a quantum
cryptographic protocol, it suffices  that the key signal is encoded
into quantum states that belong to
incompatible bases, as in the original qubit protocol of Bennett and Brassard
known as BB84~\cite{BB84}. In past years, several
qutrit-based cryptographic protocols were shown to be more secure than their
qubit-based
counterparts~\cite{Bechmann00,Bourennane01,Cerf02b,Bourennane02,Bruss02,Kaszlikowski02,Durt03}.
This justifies the
interest for studying the security of quartit protocols (see also Ref.~\cite{Bechmann99b}). In order to
evaluate the security of
such protocols against individual attacks (where Eve monitors the quartits
separately or
incoherently) we will consider a fairly general class of eavesdropping
attacks that are based
on (state-dependent) quantum cloning machines~\cite{Cerf98,Cerf00,Cerf00b}.
This will yield an
upper bound on the acceptable error rate. Higher error rates do not allow
secure
communication, since in accordance with C-K theorem~\cite{Csiszar78}, a spy
could in theory acquire all the
information, if Alice and Bob restrict themselves to one way communication
on the classical channel.

Quantum cloning is a concept that was first introduced in a seminal paper
by Buzek
and Hillery~\cite{Buzek96}, where a universal (or state-independent) and
symmetric cloning transformation was introduced for qubits. This transformation
was later extended to higher-dimensional systems by
Werner~\cite{Werner98}
but only in the special case of a universal (state-independent) cloner. In
contrast,
we will focus  on non-universal (or state-dependent) cloners.
Our starting point, summarized in Sec.~\ref{sec:state dependent cloner}
is a general characterization of asymmetric and
state-dependent
cloning transformations for $N$-level systems, as described
in Refs.~\cite{Cerf00,Cerf00b}.
In Sec.~\ref{sectiontwo} we will establish the generality  of this
formalism. We adapt the formalism in Sec.~\ref{sec:covariant cloning} such that it
is invariant  under certain unitary
transformations. We propose in Sec.~\ref{seccinq} an optical implementation
for quantum key distribution that
generalizes BB84 protocol~\cite{BB84}. In this new protocol, the signal is
encoded in
four-dimensional basis states, in two mutually unbiased bases\footnote{By
assumption, two orthonormal bases of an
N-dimensional Hilbert space are said
to be mutually unbiased if the norm of the scalar product between any two
vectors belonging each to one of the
bases is equal to ${1\over \sqrt N}$.}. The covariant cloning formalism
developed in the previous sections will be applied in
Sec.~\ref{secsix} to clone these two mutually unbiased bases that are
actually related by a double Hadamard transformation~\cite{Bechmann99b}. We will show that the
optimal cloner requires a generalization of the Bell states well adapted to
the particular problem under study,
before we conclude in Sec.~\ref{sec:conclusion}.

\section{State dependent cloning formalism \label{sec:state dependent cloner}}

To explain the  general cloning formalism it is convenient to look at
the exchange of a quantum key in a different way. Suppose Alice and
Bob exchange the maximally entangled state:
\begin{equation}
\ket{B_{00}}=\frac{1}{\sqrt{N}}\sum_{k=0}^{N-1} \ket{k}\oplus \ket{k}
\label{eq:B00 begin}
\end{equation}
The states $\ket{k}$ form a basis ($\{\ket{0},\ket{1},...,\ket{N}\}$)
(called the computational basis) in
the N-dimensional Hilbertspace. Alice has access
to the first quantum, while Bob has access to the second. If Alice now
projects her part of the state on $\ket{\psi^*}$, the resulting state is:
\begin{align}
&(\proj{\psi^*} \otimes \hat I) \sum_{k=0}^{N-1}\ket{k}\otimes \ket{k} =
 \ket{\psi^*} \sum_{i=0}^{N-1}\braket{\psi^*}{k}\otimes
 \ket{k} \nonumber\\
&=\ket{\psi^*}\sum_{i=0}^{N-1}\braket{k}{\psi}\otimes \ket{k}=\ket{\psi^*}
 \otimes \ket{\psi}
\label{eq:proj on psi*}
\end{align}
The result is a product state, and Bob has access to
$\ket{\psi}$. We see that it makes no difference
whether Alice creates and sends the state $\ket{\psi}$ to Bob, or whether
Alice projects
her part of
a shared maximally entangled state  on $\ket{\psi^*}$.
Therefore, we will use this representation of the exchange of
$\ket{\psi}$, as it makes the mathematical description of cloners more
elegant.
A spy's attack would now consist of modifying the maximally entangled state when it
was created or exchanged.

In the cloning formalism of N. Cerf~\cite{Cerf00,Cerf00b,Cerf98}, the
maximally entangled state is modified into the $N^4$ dimensional state
$\ket{\Psi}_{\text{R,A,B,C}}$. The index `R' denotes the reference
state accessible to Alice. She will project it onto $\ket{\psi^*}$ to
communicate with Bob (as she doesn't know Eve has tampered with
the maximally entangled state, she thinks that Bob will receive $\ket{\psi}$). The
index `A' denotes the state that is accessible to Bob, and it will
contain the first (imperfect) clone of $\ket{\psi}$. The index `B'
denotes the state of Eve that will contain the second clone. And
finally `C' is the state of the Cloning machine, and is also accessible
by Eve. In the formalism of N. Cerf,
$\ket{\Psi}_{R,A,B,C}$ has the form:
\begin{equation}
\label{eq:jointstate}
\ket{\Psi}_{R,A,B,C}=\sum_{m,n=0}^{N-1} a_{m,n}
\; \ket{B_{m,n}}_{R,A}
\ket{B_{m,-n}}_{B,C} = \sum_{m,n=0}^{N-1} b_{m,n} \; \ket{B_{m,n}}_{R,B}
\ket{B_{m,-n}}_{A,C}
\end{equation}
The states $\ket{B_{m,n}}$ are generalized (Fourier) Bell states
defined as~\footnote{The sum operator inside the kets is always defined
modulo N}:
\begin{equation}
\ket{B_{m,n}}_{R,A}=N^{-1/2} \sum_{k=0}^{N-1} e^{2\pi i (kn/N)}
\ket{k}_R\ket{k+m}_A
\label{bell}
\end{equation}
In~(\ref{eq:jointstate}), $a_{m,n}$ and $b_{m,n}$ are normalized
complex amplitudes:
\begin{align}
&\sum_{m,n=0}^{N-1} |a_{m,n}|^2=\sum_{m,n=0}^{N-1} |b_{m,n}|^2=1
\end{align}
We can easily calculate the inproducts of the basis states:
\begin{align}
\left(\bra{B_{m,n}}_{R,A}\,\,
\bra{B_{m,-n}}_{B,C}\right)\left(\ket{B_{x,y}}_{R,B}\,\,
\ket{B_{x,-y}}_{A,C}\right)
=N^{-1} \exp\left(\tfrac{2 \pi i}{N}(nx-my)\right)
\label{eq:inproduct of fourier bell state}
\end{align}
Using~(\ref{eq:inproduct of fourier bell state})
and~(\ref{eq:jointstate}), we can easily show that $a_{m,n}$ and
$b_{m,n}$ are dual under a Fourier transform:
\begin{align}
a_{m,n} = \frac{1}{N} \sum_{x,y=0}^{N-1} e^{2\pi i (nx-my)/ N } b_{x,y}\\
b_{m,n} = \frac{1}{N} \sum_{x,y=0}^{N-1} e^{2\pi i (nx-my)/ N } a_{x,y}
\label{eq:relation a_mn b_mn}
\end{align}

Projecting the reference system of $\ket{\Psi}_{R,A,B,C}$ onto
$\bra{\psi^*}$ yields (using (\ref{eq:jointstate}):
\begin{equation}
\sum_{m,n=0}^{N-1} a_{m,n} \; U_{m,n}\ket{\psi}_A \ket{B_{m,-n}}_{B,C}
= \sum_{m,n=0}^{N-1} b_{m,n} \; U_{m,n}\ket{\psi}_B \ket{B_{m,-n}}_{A,C}
\label{eq:state resend by Eve}
\end{equation}
with:
\begin{equation}
U_{m,n}=\sum_{k=0}^{N-1} e^{2\pi i (kn/N)} \ket{k+m}\bra{k}
\end{equation}
The error operators  $U_{m,n}$ shifts the state by $m$ units
(modulo $N$) in the computational basis, and multiplies it by a phase
to shift its Fourier transform by $n$ units (modulo $N$). Of
course, $U_{0,0}=I$, which corresponds to no error. Note that:
\begin{equation}
 (I\otimes U_{m,n})\ket{B_{0,0}}=\ket{B_{m,n}}
\end{equation}

The state~(\ref{eq:state resend by Eve}) completely characterizes the
cloning transformation\footnote{This is actually  the state that
  Eve would prepare if she intercepted a state $\ket{\psi}$ sent by
  Alice. This indeed shows that  sending $\ket{\psi}$ to Bob, or using
  the maximally entangled state, is equivalent for what concerns cloning.}.  To
evaluate the quality of the clone A (B), we take the
partial trace of over `B' and `C' (`A' and `C') of the density
operator associated with state~(\ref{eq:state resend by Eve}). The
resulting reduced density operator $\rho_A$ and $\rho_B$ of
respectively the first and the second clone are:
\begin{align}
\rho_A&=\sum_{m,n=0}^{N-1} |a_{m,n}|^2 \proj{\psi_{m,n}} \\
\rho_B&=\sum_{m,n=0}^{N-1} |b_{m,n}|^2 \proj{\psi_{m,n}}
\end{align}
where
\begin{equation}
\ket{\psi_{m,n}}=U_{m,n}\ket{\psi}
\end{equation}
The fidelity of the first clone when copying $\ket{\psi}$ (i.e the
probability that there is a collapse onto this state) is equal to
\begin{equation}\label{eq:F_A general}
F_A=\langle\psi|\rho_A|\psi\rangle
= \sum_{m,n=0}^{N-1} |a_{m,n}|^2  |\langle\psi|\psi_{m,n}\rangle|^2
\end{equation}

Focusing on the computational basis, we see that the
fidelity of any state $\ket{k}$ is equal to:
\begin{equation}   \label{eq:F_A in computational basis}
F_A=\sum_{n=0}^{N-1} |a_{0,n}|^2
\end{equation}
We also define $N-1$ disturbances: we call $D_i$ the probability that there is
a collapse on $\ket{k+i}$ if state $\ket{k}$ was sent. The
disturbances are (in the computational basis):
\begin{equation}
D_i=\sum_{n=0}^{N-1} |a_{i,n}|^2 \qquad\text{with } i\in \{1,2, \dots N-1\}
\label{eq:disturbances in the computational basis}
\end{equation}
The fidelities and the disturbances of the second clone are obtained by
replacing $a_{m,n}$ with $b_{m,n}$ in~(\ref{eq:F_A general})~--~~(\ref{eq:disturbances in the computational basis}).
It is clear that the quality of the two clones depends on the
amplitudes $a_{m,n}$ and $b_{m,n}$. Furthermore, there is  a
trade-off between the quality of the two clones. Suppose $a_{m,n}$ is
a peaked function (i.e $a_{m,n}$ is large for one value of $\{m,n\}$
and small for the other values) leading to a high-fidelity of the
first clone. As $b_{m,n}$ is the Fourier transform
of  $a_{m,n}$ it will be a rather flat function (i.e. all $b_{m,n}$
will be almost equally large), leading to a
rather low fidelity for the second clone.\footnote{Consider the extreme
case that
  $a_{0,0}=1$ and the other $a_{m,n}=0$. The first clone would in this
  case be perfect, with a fidelity of 1 and no disturbances
  ($D_i=0$). The second clone would, however, be characterized by
  $b_{m,n}=\tfrac{1}{N}$ for all values of $m$ and $n$, leading to
  fidelities and disturbances $F=D_i=\tfrac{1}{N}$.
This basically means that all values have an equal probability to be
measured, independent of the value that was sent, making the clone
worthless, as the transinformation between Alice and Bob is indeed 0.}
The balance between the quality of clones $A$ and $B$ can
be expressed, in full generality, by an entropic no-cloning
uncertainty relation that relates the probability distributions
$p_{m,n}$ and $q_{m,n}$~\cite{Cerf00b}:
\begin{equation}
H[p]+H[q] \ge \log_2(N^2)
\end{equation}
where $H[p]$ and $H[q]$ denote the Shannon entropy
of the discrete probability
distributions $p$ and $q$ defined as
\begin{align}
p(m,n)=|a_{m,n}|^2\\
q(m,n)=|b_{m,n}|^2
\end{align}
This inequality is actually a special case of a no-cloning uncertainty relation
involving the  losses of the channels that yield the two
clones~\cite{Cerf99}.
More refined uncertainty relations can be found that express
the fact that the index $m$ of output $A$ is dual to the index $n$
of output $B$, and vice versa~\cite{Cerf00b}.

\section{About the generality of Cerf's formalism\label{sectiontwo}}
One could  object that the class of cloning machines considered in this
paper is not the most
general one. Indeed, we postulate from the beginning that the joint
state $\ket{\Psi}_{R,A,B,C}$  has the form of
Eq.~(\ref{eq:jointstate}) implying
that the reduced density matrices of Alice and Bob ($R,A$) and Eve ($B,C$)
are diagonal in the Bell
basis. Although Eve is free to choose the basis that diagonalises her
reduced density matrix,  the most general state $\ket{\Psi}_{R,A,B,C}$
will not possess
the property that $Tr_{B,C}\ket{\Psi}_{R,A,B,C}\bra{\Psi}_{R,A,B,C}$ is
diagonal in the Bell basis
$\ket{B_{m,n}}_{R,A}$. We will call the  states that have
 this property ``Cerf''-states.
Nearly all the (optimal) cloning machines that
appeared in the literature can
be  unambiguously represented by a
Cerf-state~\cite{Bechmann99,Fuchs97,Bruss00,Kaszlikowski02,Cerf02c,Durt03}.
We can guess the underlying physical reason  if
we note that the Bell states are invariant (up to a global phase)
 under a cyclic relabeling of the basis states. Indeed,
 let us consider the generator $ C$  of cyclic permutations
 of the indices $ l$  of the computational basis:
\begin{equation}
C.\ket{l} = \ket{(l+1)\text{ mod } 4}
\label{eq:definition of C}
\end{equation}
 It is easy to check that the Bell
 states defined in Eq.~(\ref{bell}) are mapped onto
 themselves under such a permutation (up to a global phase),
 and under all its powers. So, when cyclic permutations are a
 natural symmetry of the protocol,
 it is also natural that the Bell states defined in Eq.~(\ref{bell}) play a
 privileged role in Eve's
 attack: the symmetries of the effects reflect the symmetries of the causes
 as is well known in physics.

Now, we could ask the inverse question:
``What is the most general state that is invariant under any cyclic
relabeling of the basis states?''
We shall show that when optimal qubit
 cloning machines are described by a $ 2^4$  dimensional
 pure state which
 is invariant under any cyclic relabeling of the basis states,
 such a state is a Cerf state. This shows (at least for qubits)
 that Cerf's formalism for optimal cloning machines is more general
 than it could seem at first sight.

In two dimensions, only two permutations exist: the identity
 and the generator $ C$ that switches the labels 0 and 1 (which is a parity
operator: $C^2$ = $1$). We
 shall now prove the following theorem:

{ \bf Theorem}:

Let us consider any qubit protocol. Let us consider the corresponding 16
dimensional pure state that is assumed
to be optimal and to be invariant under any cyclic relabeling of the basis
states. Such a state is a Cerf-state
so to say $Tr_{B,C}\ket{\Psi}_{R,A,B,C}\bra{\Psi}_{R,A,B,C}$ is
diagonal in the Bell basis
$\ket{B_{m,n}}_{R,A}$.

{ \bf Proof}:

The four Bell states, when $N$ = 2, are defined as follows:
\beqa
&&\ket{B_{0,0}}=\frac{1}{\sqrt{2}}(\ket{00}+\ket{11})\qquad
\ket{B_{1,0}}=\frac{1}{\sqrt{2}}(\ket{01}+\ket{10})\\
&&\ket{B_{0,1}}=\frac{1}{\sqrt{2}}(\ket{00}-\ket{11})\qquad
\ket{B_{1,1}}=\frac{1}{\sqrt{2}}(\ket{01}-\ket{10})
\eeqa
 It is easy to check that the Bell states
 are eigenstates of the permutation
operator $ C$  with the eigenvalues $\pm1$. If we impose that the joint
state $ \ket{\Psi}_{R,A,B,C}$  is invariant under the action of the
 permutation operator ($ C \ket{\Psi}_{R,A,B,C}$
  = $ \ket{\Psi}_{R,A,B,C}$
 ), then it belongs
necessarily to the 8 dimensional eigenspace associated to the
 eigenvalue +1 of $ C$:
\begin{align}
\ket{\Psi}_{R,A,B,C}=\,\,&\alpha_+\ket{B_{0,0}}_{R,A}\otimes\ket{B_{0,0}}_{B,C}+
\alpha_-\ket{B_{0,1}}_{R,A}\otimes\ket{B_{0,1}}_{B,C}\nonumber\\
+&\beta_+\ket{B_{1,0}}_{R,A}\otimes\ket{B_{1,0}}_{B,C}+
\beta_-\ket{B_{1,1}}_{R,A}\otimes\ket{B_{1,1}}_{B,C}\nonumber\\
+&\gamma_+\ket{B_{1,0}}_{R,A}\otimes\ket{B_{0,0}}_{B,C}+
\gamma_-\ket{B_{1,1}}_{R,A}\otimes\ket{B_{0,1}}_{B,C}\nonumber\\
+&\delta_+\ket{B_{0,0}}_{R,A}\otimes\ket{B_{1,0}}_{B,C}+
\delta_-\ket{B_{0,1}}_{R,A}\otimes\ket{B_{1,1}}_{B,C}
\label{eq:general state}
\end{align}
We see that the requirement that the full state is invariant under a
permutation of the basis states reduces the
generality of the joint state (which is usually described by 16
amplitudes). Here the amplitudes associated to
states that are antisymmetric under the action of the permutation operator
$C$ are assumed to be equal to zero. Although Eq.~(\ref{eq:general state})
does not yet describe a
Cerf-state, we shall proof that in the optimal case it does. In order to do
so, let
us first
consider the reduced state
shared by Alice and Bob:
\begin{align}
\rho_{R,A}=&Tr_{B,C}\ket{\Psi}_{R,A,B,C}\bra{\Psi}_{R,A,B,C}\nonumber\\
=&(\alpha_+\ket{B_{0,0}}_{R,A}+\gamma_+\ket{B_{1,0}}_{R,A})
(\alpha^*_+\bra{B_{0,0}}_{R,A}+\gamma^*_+\bra{B_{1,0}}_{R,A})\nonumber\\
+&(\beta_+\ket{B_{1,0}}_{R,A}+\delta_+\ket{B_{0,0}}_{R,A})
(\beta^*_+\bra{B_{1,0}}_{R,A}+\delta^*_+\bra{B_{0,0}}_{R,A})\nonumber\\
+&(\alpha_-\ket{B_{0,1}}_{R,A}+\gamma_-\ket{B_{1,1}}_{R,A})
(\alpha^*_-\bra{B_{0,1}}_{R,A}+\gamma^*_-\bra{B_{1,1}}_{R,A})\nonumber\\
+&(\beta_-\ket{B_{1,1}}_{R,A}+\delta_-\ket{B_{0,1}}_{R,A})
(\beta^*_-\bra{B_{1,1}}_{R,A}+\delta^*_-\bra{B_{0,1}}_{R,A})
\end{align}
It is worth noting that the non-diagonal components of the reduced density
matrix $\rho_{R,A}$ of the
type $\ket{B_{i,j}}\bra{B_{m,n}}$ (for $i\not= m$) do not
contribute  to the
statistics of the  measurements performed by Alice and Bob in the
computational basis. Indeed:

\begin{align}
\braket{pq}{B_{i,j}} \braket{B_{m,n}}{pq} &=
 \frac{1}{2} \sum_{k,l} (-1)^{kj+ln} \braket{p}{k} \braket{q}{k+i}
 \braket{l}{p} \braket{l+m}{q}\\
 &=\frac{1}{2} \sum_{k,l} (-1)^{kj+ln} \delta_{p,k} \; \delta_{q,k+i} \;
 \delta_{l,p}\; \delta_{l+m,q}\\
 &=\frac{1}{2} (-1)^{p(j+n)} \delta_{q,p+i}\; \delta_{q,p+m}
\end{align}
which is $0$ if $m\ne i$. This means that everything happened, from the
point of view of
Alice and Bob, as if
$\rho_{R,A}$ was equal to the effective density operator
$\rho^{eff}_{R,A}$ defined as:
\begin{align}
\rho^{eff}_{R,A}=&\nonumber
(|\alpha_+|^2+|\delta_+|^2)\ket{B_{0,0}}\bra{B_{0,0}}+
(|\beta_+|^2+|\gamma_+|^2)\ket{B_{1,0}}\bra{B_{1,0}}\nonumber \\
&+(|\alpha_-|^2+|\delta_-|^2)\ket{B_{0,1}}\bra{B_{0,1}}+
(|\beta_-|^2+|\gamma_-|^2)\ket{B_{1,1}}\bra{B_{1,1}}
\label{asif}
\end{align}
 Let us evaluate the information possessed by Eve relatively to
Alice. We denote
$P^E_{i,j}$ the probability that Eve simultaneously observes the output
clone $B$ to be in the $i^\text{th}$ detector and the cloning machine
$C$ in the $j^\text{th}$ detector and $P^{A|E}_{k;i,j}$ the probability
that Alice's
$k^\text{th}$ detector
fires while Eve simultaneously observes the output
clone $B$ in the $i^\text{th}$
detector and the cloning machine
$C$ in the $j^\text{th}$ detector. By direct computation, we obtain that:
\begin{align}
P^E_{0,0}=&P^E_{1,1}=\frac{1}{2}(|\alpha_+|^2+|\alpha_-|^2+|\gamma_+|^2+
|\gamma_-|^2) \label{eq:P00}\\
P^E_{0,1}=&P^E_{1,0}=
\frac{1}{2}(|\beta_+|^2+|\beta_-|^2+|\delta_+|^2+|\delta_-|^2)\\
P^{A;E}_{0;0,0}=&1-P^{A;E}_{1;0,0}=1-P^{A;E}_{0;1,1}=P^{A;E}_{1;1,1}\nonumber\\
=&\frac{1}{2}P^E_{0,0}+\frac{1}{2}\text{Re}(\alpha_+\cdot \alpha_-^*)
+\frac{1}{2}\text{Re}(\gamma_+\cdot \gamma_-^*)
\label{stati}\\
P^{A;E}_{0;0,1}=&1-P^{A;E}_{1;0,1}=1-P^{A;E}_{0;1,0}=
P^{A;E}_{1;1,0}\nonumber\\
=& \frac{1}{2} P^E_{0,1} +\frac{1}{2}\text{Re}(\beta_+\cdot \beta_-^*)
+\frac{1}{2}\text{Re}(\delta_+\cdot \delta_-^*)  \label{stati2}
\end{align}
Equations~(\ref{eq:P00}--\ref{stati2}) show that the information gained by Eve
during such an attack is
certainly not superior to the information gained during an attack in which
Eve chooses the Cerf state
\begin{align}
\frac{1}{\sqrt{P_1}}
      &\left( \alpha_+\ket{B_{0,0}}_{R,A}\otimes\ket{B_{0,0}}_{B,C}+
\alpha_-\ket{B_{0,1}}_{R,A}\otimes\ket{B_{0,1}}_{B,C}   \right.     \nonumber\\
& \left. +\beta_+\ket{B_{1,0}}_{R,A}\otimes\ket{B_{1,0}}_{B,C}
+\beta_-\ket{B_{1,1}}_{R,A}\otimes\ket{B_{1,1}}_{B,C} \right)
\label{eq:half attack 1}
\end{align}

 with probability
$P_1=|\alpha_+|^2+|\alpha_-|^2+ |\beta_+|^2+|\beta_-|^2$ and the Cerf state
\begin{align}
\frac{1}{\sqrt{P_2}} & \left(
\gamma_+\ket{B_{0,0}}_{R,A}\otimes\ket{B_{0,0}}_{B,C}+
\gamma_-\ket{B_{0,1}}_{R,A}\otimes\ket{B_{0,1}}_{B,C} \right) \nonumber\\
&\left.+ \delta_+\ket{B_{1,0}}_{R,A}\otimes\ket{B_{1,0}}_{B,C}+
\delta_-\ket{B_{1,1}}_{R,A}\otimes\ket{B_{1,1}}_{B,C}\right)
\label{eq:half attack 2}
\end{align}
 with probability
$P_2= |\gamma_+|^2+|\gamma_-|^2 +|\delta_+|^2+|\delta_-|^2$ .
Furthermore, the two attacks lead to the same distribution of
statistical results, as the effective density operator~(\ref{asif})
equals  the density operator of the attack using~(\ref{eq:half
  attack 1}) and~(\ref{eq:half attack 2}).
By assumption,
the optimal (pure) Cerf state is superior to this second attack (which uses a
mixture of Cerf states), which ends the proof.
 The generalization of this
proof to arbitrary dimensions is given in~\cite{Durt03b}.

\section{The covariant cloning machine\label{sec:covariant cloning}}

The formalism defined in the previous section is valid when Alice and Bob
respectively encode
and measure the signal in the computational basis
$(\ket{0},\ket{1},\ket{2}...\ket{N})$. Quantum cryptographic
protocols impose that Alice must use at least another basis $\tilde\psi$
$(\ket{\tilde\psi_0},\ket{\tilde\psi_1},\ket{\tilde\psi_2},\dots
\ket{\tilde \psi_N})$, with
\begin{equation}
\braket{i}{\tilde\psi_j} = A_{ ij}
\end{equation}
 If Alice and Bob share the joint
state
$\ket{B_{0,0}}$ and that Alice wishes to encode the signal in the
$\tilde\psi$ basis, she must project her
component of $\ket{B_{0,0}}$ into the conjugate basis (that we shall from
now on denote the $\tilde\psi^*$ basis)
defined as follows: $\braket{i}{\tilde\psi_j^*}$ =
$A^*_{ ij}$. This
property is a direct consequence of Eq.~(\ref{eq:proj on psi*}).
We now require that the
computational basis is not privileged, and
that the cloning machine clones the states of the computational and
the $\tilde\psi$ basis in the same manner. This implies
that after projecting the reference (Alice's) system onto the state
$\ket{\tilde\psi^*}$ (which would result in an input state
$\ket{\tilde\psi}$ of Bob, in the absence of an eavesdropper), the  reduced
density
operators  of Bob and Eve must be of
the form:
\begin{align} \label{rho}
&\rho_A=\sum_{m,n=0}^{N-1} p_{m,n} \proj{\tilde\psi_{m,n}} \\
&\rho_B=\sum_{m,n=0}^{N-1} q_{m,n} \proj{\tilde\psi_{m,n}}\\
\intertext{where}
&\ket{\tilde\psi_{m,n}}=\tilde U_{m,n}\ket{\tilde\psi}\\
\intertext{and}
&\tilde U_{m,n}=\sum_{k=0}^{N-1} e^{2\pi i (kn/N)}
\ket{\tilde\psi_{k+m}} \bra{\tilde\psi_k}.
\end{align}
A possible corresponding joint state
of the two clones and the cloning machine is (see also~\cite{Durt03}):
\beq
\sum_{m,n=0}^{N-1} a_{m,n} \; \tilde U_{m,n}\ket{\tilde \psi}_A \ket{\tilde
B_{m,n}^*}_{B,C}
= \sum_{m,n=0}^{N-1} b_{m,n} \; \tilde U_{m,n}\ket{\tilde \psi}_B
\ket{\tilde B_{m,n}^*}_{A,C}
\label{eq:joint state na proj psi}
\eeq
where
\beq
\label{belltilde*}\ket{\tilde B_{m,n}}=N^{-1/2} \sum_{k=0}^{N-1} e^{2\pi
i (kn/N)}
\ket{\tilde\psi_k^*}\ket{\tilde\psi_{k+m}}=(I\otimes \tilde U_{m,n})
\ket{B_{0,0}}
\eeq

Equation~(\ref{belltilde*}) is a  generalization of the Bell States
defined in~(\ref{bell}). Indeed, if $\ket{\tilde \psi}$ is the
computational basis,~(\ref{belltilde*}) and~(\ref{bell}) are equal, because
the computational basis is real and $\ket{ B_{m,n}^*}=\ket{ B_{m,-n}}$.
The joint state of the reference $R$, the
two output clones ($A$ and $B$), and the ($N$-dimensional) cloning machine
$C$ that corresponds with~(\ref{eq:joint state na proj psi}) is:
\begin{equation}
\sum_{m,n=0}^{N-1} a_{m,n} \; \ket{\tilde B_{m,n}}_{R,A} \ket{\tilde
B_{m,n}^*}_{B,C}
= \sum_{m,n=0}^{N-1} b_{m,n} \; \ket{\tilde B_{m,n}}_{R,B} \ket{\tilde
B_{m,n}^*}_{A,C}
\end{equation}

With this choice\footnote{This choice is unique up to arbitrary unitary
transformations in
the
$N^2$ dimensional Hilbert space assigned to Eve ($B$ and $C$)}, the
requirement of covariance
in the computational and the $\tilde\psi$ bases imposes the extra-constraint:
\begin{equation} \label{cov}
\sum_{m,n=0}^{N-1} a_{m,n} \; \ket{B_{m,n}}_{R,A} \ket{B_{m,n}^*}_{B,C} =
\sum_{i,j=0}^{N-1} a_{i,j}
\; \ket{\tilde B_{i,j}}_{R,A} \ket{\tilde B_{i,j}^*}_{B,C}
\end{equation}
As the $B_{m,n}$ states form an orthonormal basis, we can write:
\begin{align}
&\sum_{m,n=0}^{N-1} a_{m,n} \ket{B_{m,n}}_{R,A} \ket{B_{m,n}^*}_{B,C}
\nonumber\\
=&\sum_{i,j} a_{i,j}
\left(\sum_{m,n}\ketbra{B_{m,n}}{B_{m,n}}\right)
 \ket{\tilde  B_{i,j}}_{R,A}
 \left( \sum_{k,l} \ketbra{B_{k,l}^*}{B_{k,l}^*}\right)
\ket{\tilde B_{i,j}^*}_{B,C}
\end{align}
Defining:
\begin{align}
V_{i,j,k,l}\equiv \braket{B_{i,j}}{\tilde  B_{k,l}}
\end{align}
we get:
\begin{align}
&\sum_{\substack{m,n\\k,l}} a_{m,n} \delta_{(m,n),(k,l)}\ket{B_{m,n}}_{R,A}
\ket{B_{k,l}^*}_{B,C}\\
&=
\sum_{\substack{i,j,k\\l,m,n}} a_{i,j}
 \ket{B_{m,n}}_{R,A}  V_{m,n,i,j}
\ket{B_{k,l}^*}_{B,C} V^*_{k,l,i,j}\\
&=\sum_{\substack{m,n,i,p\\j,k,l,q}} a_{i,j} \delta_{(i,j),(p,q)} V_{m,n,i,j}
V^*_{k,l,p,q}  \ket{B_{m,n}}_{R,A}  \ket{B_{k,l}^*}_{B,C}
\end{align}

Formally, this constraint can be expressed as a matrix relation:
\begin{equation}
 {\cal V}{\cal A}={\cal A}{\cal V}
\label{eq:commutation relation}
\end{equation}
 where ${\cal V}$ and ${\cal A}$ are $N^2$x$N^2$
matrices defined as
follows:
\begin{align}
&{\cal V}_{m,n;i,j} =V_{m,n,i,j}
\\\intertext{and ${\cal A}$ the diagonal matrix defined as}
&{\cal A}_{i,j;p,q}= a_{i,j}\delta_{(i,j),(p,q)}
\end{align}
 As ${\cal A}$ is diagonal, (\ref{eq:commutation relation}) is
 extremely simple to solve:
\begin{equation}
\text{if } {\cal
V}_{i,j;p,q}\not = 0 \quad \text{ then }
a_{i,j} = a_{p,q}
\label{eq:condition covariant a_mn}
\end{equation}
 Therefore, to build a cloning machine that
is invariant in two
bases (the computational basis and the $\tilde\psi$ basis) we only
need to  compute
the $N^4$ inproducts ${\cal V}_{i,j;k,l} = \braket{B_{i,j}}{\tilde
B_{k,l}}$. The solutions $a_{m,n}$ of Eq.~(\ref{eq:condition covariant
a_mn}) define the most
general cloning machine
that is invariant in the two bases.

The generalization in the case where none of the bases
is the computational basis is straightforward. The condition
(\ref{eq:condition covariant a_mn}) remains valid, as long as we
construct the Bell states using~(\ref{belltilde*}) for the two bases.

\section{\label{seccinq} Quantum cryptography with two complementary quartit
bases: A protocol based on interferometric complementarity.}
\begin{figure}
\includegraphics[width=5cm]{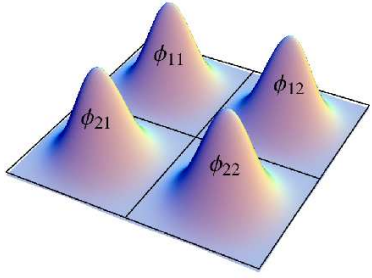}
\caption{\label{fig:rectangarray} Rectangular Array of pulse shapes}
\end{figure}

A quantum protocol for  key distribution that implements quartit
states can be conceived using the lateral shape of
electromagnetic pulses. Consider figure~\ref{fig:rectangarray}.
Each $\phi_{ij}$ represents the same (lateral) field distribution, shifted
in its own
square, and coherent with respect to the others. In theory any
distribution could be used, as long as two neighboring distributions do not
overlap. In practice, these can be easily produced by a fan-out of a
single laser source. In combination  with an external modulator, that
can switch off or switch on each channel, and a controllable $\pi$-phase
retarder, we can easily realize the following two mutually unbiased bases:
\beqa
\ket{0}&=\frac{1}{\sqrt{2}} \left(\ket{\phi_{11}} +
  \ket{\phi_{12}}\right) \label{eq:ket0}
&\ket{1} =\frac{1}{\sqrt{2}} \left(\ket{\phi_{11}} -
  \ket{\phi_{12}}\right)\nonumber\\
\ket{2}&=\frac{1}{\sqrt{2}} \left(\ket{\phi_{21}} +
  \ket{\phi_{22}}\right)
&\ket{3}=\frac{1}{\sqrt{2}} \left(\ket{\phi_{21}} -
  \ket{\phi_{22}}\right)\label{base}\eeqa
and
\beqa \ket{0'}&=\frac{1}{\sqrt{2}} \left(\ket{\phi_{11}} +
  \ket{\phi_{21}}\right)
&\ket{1'}=\frac{1}{\sqrt{2}} \left(\ket{\phi_{11}} -
  \ket{\phi_{21}}\right)\nonumber\\
\ket{2'}&=\frac{1}{\sqrt{2}} \left(\ket{\phi_{12}} +
  \ket{\phi_{22}}\right)
&\ket{3'}=\frac{1}{\sqrt{2}} \left(\ket{\phi_{12}} -
  \ket{\phi_{22}}\right) \label{base'}
\eeqa
so we have :
\begin{equation}
|\braket{i}{j'}|^2=\frac{1}{4}
\end{equation}
If we represent the in-products between the first basis and the second one
in a matrix form, we get:
\begin{equation}\label{adam}
H_{ij}=(\braket{i}{j'})= \frac{1}{2}\left(\begin{array}{cccc}
        +1&+1&+1&+1\\
        +1&+1&-1&-1\\
        +1&-1&+1&-1\\
        +1&-1&-1&+1\\
\end{array}\right)
\end{equation}
Actually, this is (up to a permutation of the second and the third primed
basis states) a double Hadamard transform~\cite{Bechmann99b,Brainis03}. The simple
Hadamard transform is well-known in
quantum information~\cite{Nielsen}: it sends the qubit
$\ket{i}$ ($i$=0, 1) onto $\frac{1}{\sqrt{2}}(\ket{0}+(-)^i \ket{1})$. It is
easy to check that if we identify the
computational quartit basis with the product basis as follows:
$\ket{0^{quart}}=\ket{00},\ket{1^{quart}}=\ket{10},\ket{2^{quart}}
=\ket{01},$ and $\ket{3^{quart}}=\ket{11}$, then, when the qubits undergo a
simple Hadamard transformation,
quartits change according to Eq.~(\ref{adam}) (up to a permutation of the
labels 2 and 3 of the primed basis).

\begin{figure}
\includegraphics[width=8cm]{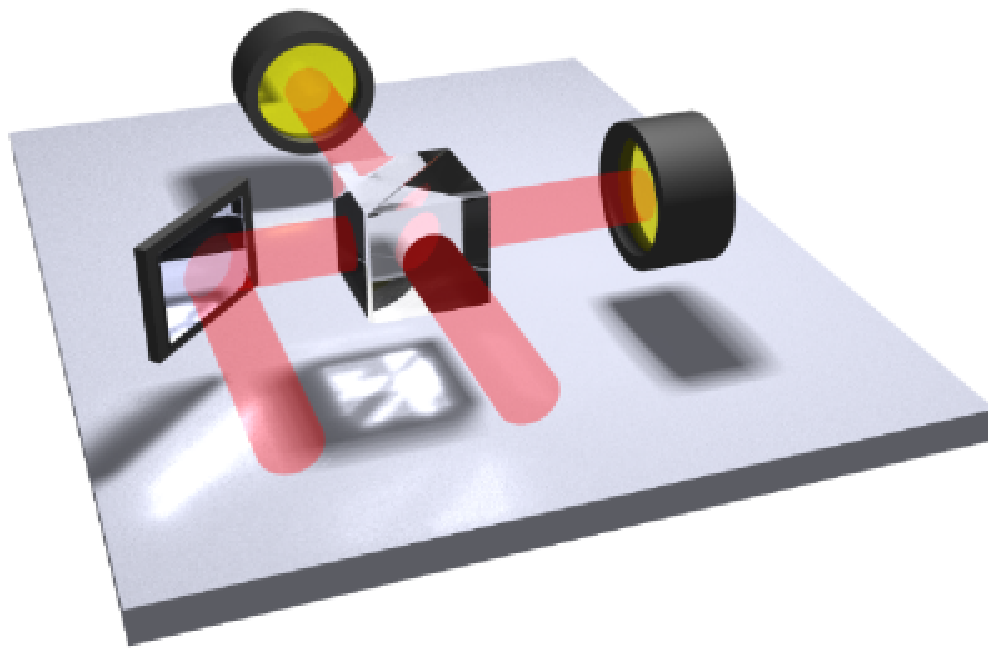}
\includegraphics[width=6cm]{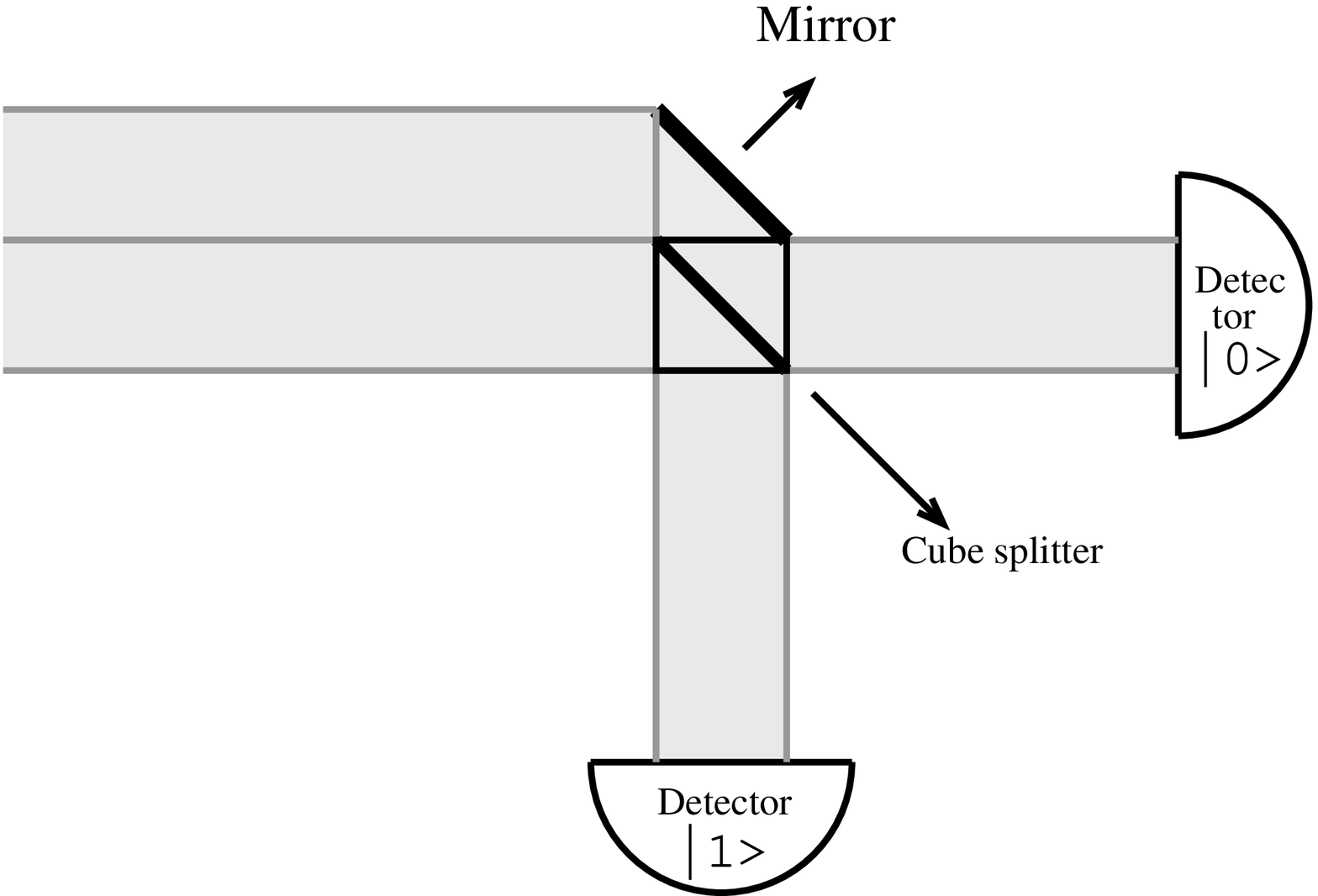}
\caption{\label{fig:measuringapp} Possible measuring apparatus,
  distinguishing between $\ket{0}$ and $\ket{1}$. The distances of the
  mirrors are tuned in such a way that in the horizontal (vertical) arm
there is
  constructive (destructive) interference }
\end{figure}

Therefore we shall from now
on call it a Hadamard transform, and denote
its elements
$H_{ij}$ in accordance with Eq.~(\ref{adam}). Figure
\ref{fig:measuringapp} shows a relatively simple setup, consisting of a
mirror, a 50/50 beam splitter and two
photon counters, which can distinguish between the states.
The mirror's position is tuned to ensure that if two distributions
enter the device, their fields are added in the horizontal branch and
subtracted in the vertical one. So if we position the two input lines
of this  interferometer in the $\phi_{11}$ and $\phi_{12}$ channel,
the photon counter in the horizontal(vertical) branch corresponds to a
projective measurement on $\ket{0}$($\ket{1}$). In a similar way, and
by rotating the setup (or the incoming channels) projections on the
$\ket{i'}$-basis can be made.

\section{Estimation of the safety threshold.\label{secsix}}
\subsection{Applying the covariant formalism to Fourier complementary bases
\label{sixa}}
\label{sec:Applying cov form to fourier}
We will now apply the covariant formalism to find a cloner that
clones equally well two quartit bases,
that are  Fourier
transforms of each other. Cloning between such bases has been studied  in
the past~\cite{Cerf02,Cerf02b}. We would like to compare the results
of our covariant cloner with the optimal cloners of the literature.  We
have the computational basis$ \{\ket{0}$, $\ket{1}$,
$\ket{2}$, $\ket{3}\}$
and its Fourier transform $\ket{k'}=\tfrac{1}{2}\sum_{j=0}^{3}
e^{ 2 \pi i (kj/4)} \ket{j}$:
\begin{align}
\ket{0'}&=\tfrac{1}{2}(\ket{0}+\ket{1}+\ket{2}+\ket{3})\nonumber\\
\ket{1'}&=\tfrac{1}{2}(\ket{0}+i\ket{1}-\ket{2}-i\ket{3})\nonumber\\
\ket{2'}&=\tfrac{1}{2}(\ket{0}-\ket{1}+\ket{2}-\ket{3})\nonumber\\
\ket{3'}&=\tfrac{1}{2}(\ket{0}-i\ket{1}-\ket{2}+i\ket{3})
\label{eq:Fourier bases}
\end{align}
The $\ket{k'}$ basis plays the role of $\ket{\tilde \psi}$
in Sec.~\ref{sec:covariant cloning}.
We denote the unitary transformation that connects the two bases as
$F$:
\begin{align}
F_{k,l}&=\braket{k}{l'}=\tfrac{1}{2} e^{ 2 \pi i( kj/4)} = \tfrac{1}{2}
i^{kl} \\
&=\frac{1}{2}\left(\begin{array}{rrrr}  1 & 1 & 1 & 1 \\
                      1 & i &-1 &-i \\
                      1 &-1 & 1 &-1 \\
                      1 &-i & -1& i
 \end{array}\right)
\label{eq:definition F qc}
\end{align}
Calculating the $16\times16$ matrix ${\cal V}_{i,j;k,l}$(we write
$\ket{\tilde k}\equiv\ket{k'}$):
\begin{align}
{\cal V}_{m,n;k,l}&=\braket{B_{m,n}}{\tilde B_{k^*,l}}\\
&=\tfrac{1}{4}\sum_{p,q=0}^{3}\exp\left(\tfrac{2 \pi i}{2}(ql-pn)\right)
\braket{p}{\tilde q^*}\braket{p+m}{\tilde{q+k}}\\
&=\tfrac{1}{16}\sum_{p,q=0}^{3} i^{(ql-pn)} i^{-pq} i^{(p+m)(q+k)}\\
&=\tfrac{1}{16}
i^{mk}\sum_{p=0}^{3}i^{(k-n)p}\sum_{q=0}^{3}i^{(l+m)q}\\
&=i^{mk}\delta_{l+m,4}\delta_{k,n}
\end{align}
Remember that all additions are modulo 4. Applying~(\ref{eq:condition
  covariant a_mn}) we get the condition:
\begin{align}
a_{m,n}=a_{k,l} \qquad \text{ if } \;\; l+m=0 \;\text{ and }\; k=n
\end{align}
leading to:
\begin{align}
&a_{01}=a_{03}=a_{10}=a_{30} & & a_{02}=a_{20}\\
&a_{21}=a_{12}=a_{23}=a_{32} & & a_{11}=a_{13}=a_{33}=a_{31}
\end{align}
This corresponds to the following $a_{mn}$ matrix:
\begin{equation}
(a_{m,n})= \left(\begin{array}{cccc}
        a&b&c&b\\
        b&d&f&d\\
        c&f&e&f\\
        b&d&f&d \\
\end{array}\right)
\end{equation}
If we desire to optimize the information of Eve, conditioned on Alice and
Bob's observations, this means
that certain conditions of constructive interference must be fulfilled.
Moreover, the disturbances must be the
same because the transmission line is assumed to be isotropic. This leads
eventually~\cite{Durt03b} to
the condition that $b=c$ and $e=f$. So the
$a_{mn}$  matrix that is covariant in the two Fourier bases becomes:
\begin{equation}
(a_{m,n})=\begin{pmatrix} x & y & y & y\\
                          y & z & z & z\\
                          y & z & z & z\\
                          y & z & z & z
          \end{pmatrix}
\label{eq:matrix a_mn in fourier basis}
\end{equation}
which leads to the cloning state:
\begin{align}
 \ket{\Psi}_{R,A,B,C}=&\sum_{m,n=0}^{3}  a_{m,n} \ket{B_{m,n}}_{R,A}
 \ket{B_{m,n}^*}_{B,C}\nonumber \\
=& (v-2x+y)\ket{B_{0,0}}_{R,A}
\ket{B_{0,0}}_{B,C}+
 (x-y)\sum_{n=0}^{3} \ket{B_{0,n}}_{R,A} \ket{B_{0,n}^*}_
 {B,C}\nonumber\\
&+(x-y)\sum_{m=0}^{3} \ket{B_{m,0}}_{R,A} \ket{B_{m,0}}_ {B,C}
+y\sum_{m=0}^{3}\ket{B_{m,n}}_{R,A}
\ket{B_{m,n}^*}_{B,C}\label{eq:cloneF}
\end{align}

Actually, it is easy to check\cite{Durt03} that, according to Eq.~
(\ref{belltilde*}),
\begin{align} \label{dual1}
&\ket{\tilde B_{m,n}}_{R,A}=i^{-nm} \ket{B_{-n,m}}_{R,A}\\
&\ket{\tilde B_{m,n}^*}_{B,C}=i^{+nm}\ket{B_{-n,m}^*}_{B,C}
\label{dual2}
\end{align}
There exists thus a bijective relation between the Bell states expressed
in the computational basis and those expressed in the Fourier
basis. It helps to
 understand the invariance of the state $ \ket{\Psi}^F_{R,A,B,C}$  in
 both bases. Thanks to this bijective relation, it is  easy to show that:
\beqa
\ket{B_{0,0}}_{R,A}\ket{B_{0,0}}_{B,C}=\ket{ \tilde B_{0,0}}_{R,A}
 \ket{ \tilde B_{0,0}^*}_{B,C}\nonumber \\
\sum_{n=0}^{3}\ket{B_ {0,n}}_{R,A} \ket{B_{0,n}^*}_{B,C}=\sum_{m=0}^{3}
\ket{\tilde B_{m,0}}_{R,A} \ket{\tilde B_{m,0}^*}_{B,C}\nonumber \\
\sum_{m=0}^{3}\ket{ B_{m,0}}_{R,A} \ket{B_{m,0}^*}_{B,C}=\sum_{m=0}^{3}
\ket{\tilde B_{0,m}^*}_{R,A} \ket{\tilde B_{0,m}^*}_{B,C}\nonumber \\
\sum_{m=0}^{3}\ket{B_{m,n}}_{R,A} \ket{B_{m,n}^*}_{B,C}=
\sum_{m=0}^{3}\ket{\tilde B_{m,n}}_{R,A} \ket{\tilde B_{m,n}^*}_{B,C}
\nonumber
\eeqa
A cloning machine with such a matrix was already studied
in~\cite{Cerf02b}, although in that case the authors proposed the
matrix without the covariant demand.\footnote{Of course this choice
  was not random. It was shown~\cite{Cerf02,Cerf02b}
  that such a matrix clones all the states that belong to
  the two Fourier bases with the same fidelity. However there is no
  guarantee that there are no other matrices which have this
  property. If we demand covariance of the cloner state, we have the
  assurance that~(\ref{eq:matrix a_mn in fourier basis}) is the only
  solution.}
The optimum of this cloner (in four dimensions) and in the symmetric
case (i.e. the cloner produces two clones of the same quality) is
characterized by~\cite{Cerf02,Cerf02b}:
\begin{equation}
(a_{m,n})=\frac{1}{4}\begin{pmatrix}
3 &        1     &        1     &     1        \\
1 & \slantfrac{1}{3} &\slantfrac{1}{3} &\slantfrac{1}{3}   \\
1 & \slantfrac{1}{3} &\slantfrac{1}{3} &\slantfrac{1}{3}   \\
1 & \slantfrac{1}{3} &\slantfrac{1}{3} &\slantfrac{1}{3}
          \end{pmatrix}
\label{eq:matrix a_mn in fourier basis numeric optimal}
\end{equation}
The two clones have a fidelity of
\begin{equation}
\label{eq:Fidelity Fourier}
F=\slantfrac{3}{4}
\end{equation}
and the three
disturbances (error rates)
\begin{equation}
D_1=D_2=D_3=\slantfrac{1}{{12}}
\end{equation}
 As the
three Disturbances are always equal with the matrix~(\ref{eq:matrix a_mn
  in fourier basis}), the maximum fidelity will also be the maximum of
the mutual information between Alice and Bob (or between  Alice and Eve
for the second clone):
\begin{equation}
I_{AB}=I_{AE}=0.792
\end{equation}
A theorem of Csisz\'{a}r and K\"{o}rner
\cite{Csiszar78}  shows that a secret key with $R$ bits can be
generated through privacy amplification if:
\begin{equation}
\label{csiszar}
R \geq I_{AB}-I_{AE}
\end{equation}
It is therefore sufficient that $I_{AB}>I_{AE}$ in order to establish a secret
key. If we restrict ourselves to one-way communication
on the classical channel, (\ref{csiszar}) is also a necessary
condition. Therefore, an error rate of 25\% is the maximum that can be
allowed between Alice and Bob.

As we noted in section \ref{sectiontwo}, the Cerf states are invariant
under any
cyclic permutation of the labels of the
 basis states. Actually, the invariance under cyclic permutations of
 the indices of the
 computational (Fourier) basis plays a fundamental role in this
 approach. For instance, it is easy to check that the Bell state
 $ \ket{B_{m,n}}_{X,Y}$  is obtained
 by projecting the state $ \ket{0,m}$
   onto the eigenspace of $ C$  associated with
 the eigenvalue $ i^{-n}$ ($C$ is defined in Eq.~(\ref{eq:definition
 of C})). This projector is
 equal to
\begin{equation}
 \sum_{k=0}^{4-1} \ket{B_{kn}}\bra{B_{kn}}
=\frac{1}{4}\sum_{k=0}^{4-1}i^{nk}C^k
\label{eq:projector in fourier case}
\end{equation}
 where we made use
 of the fact that $ C^4=C^0=1$. Beside, one can check that
 the same Bell state, when it is expressed in the product bases
 $  \ket{\tilde\psi_k}\ket{\tilde\psi_{k'^*}}$  (and $  \ket{\tilde\psi_k^*}
\ket{\tilde\psi_{k'}}$ ) is invariant under cyclic
 relabeling of the indices $ k,k'$  of the type $ k,k'\to k+1,k'+1$
 (let us call these permutations $ \tilde C$  and $ \tilde C'$),
for the eigenvalues $ i^{m}$  and $ i^{-m}$  respectively. We shall
generalize this property in a forthcoming section.

\subsection{Estimation of the safety threshold
 in the usual formalism.}Let us now study the security of the
 protocol based on interferometric complementarity using the
 covariant cloning machines described in the previous section.
 The condition (${\cal V}_{i,j;k,l}\not = 0$, then $a_{i,j}$ =
$a_{k,l}$) yields the $a_{mn}$ matrix:
\begin{equation}
(a_{m,n})= \left(\begin{array}{cccc}
        a&b&c&b\\
        b&d&b&d\\
        c&b&e&b\\
        b&d&b&d \\
\end{array}\right)
\end{equation}

The associated matrix $b_{mn}$ that we obtain according to Eq.~(\ref{eq:relation a_mn b_mn}) is:
\begin{equation}
(b_{m,n})= {1\over 4}\left(\begin{array}{cccc}
        a+8b+2c+2d+e&a-e&a-e&a-e\\
        a-e&a-2c+e&a-e&a-2c+e\\
        a+2c-4d+e&a-e&a-4b+2c+e&a-e\\
        a-e&a-2c+e&a-e&a-2c+e \\
\end{array}\right)
\end{equation}

\begin{figure}
\includegraphics[width=8cm]{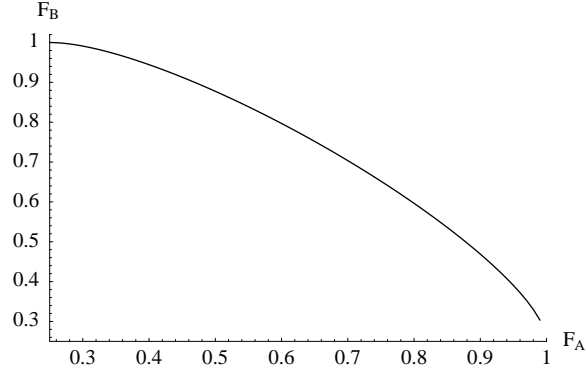}
\caption{\label{fig:fidelities} Fidelity of the second clone as a
  function of the fidelity of the first clone. They equalize at
  $F=0.7018$.}
\end{figure}

According to Eq.~(\ref{eq:F_A in computational basis}),
 one can check that the fidelity associated to such a cloning
 state is equal to $ F_A=|a|^2+2|b| ^2+|c|^2$.
Let us define the disturbance $ D^i_A$  ($ i=1,2,3$ )
 as the probability that when Alice measures the state $ \ket{j}$ , Bob
 measures the state $ \ket{j+i}$. It is easy to check that
\beq \label{disturb}
D^i_A= \sum_{n=0}^{N-1} |a_{i,n}|^2
\eeq
As the transmission line is usually considered to be isotropic,
 we have   $ D^1_A=D^2_A=D^3_A$, yielding the condition that $
2|d|^2=|c|^2+|e|^2$. The cloning state defined by the
 following matrix $a^{iso}_{mn}$ obviously fulfills these constraints:
\begin{equation}
(a^{iso}_{m,n})= \left(\begin{array}{cccc}
        a&b&c&b\\
        b&c&b&c\\
        c&b&c&b\\
        b&c&b&c \\
\end{array}\right)
\end{equation}
The fidelity and disturbances are:
\begin{align}
F^{iso}_A =|a|^2 +2|b|^2 + |c|^2 \\
D^{iso1}_A=D^{iso2}_A=D^{iso3}_A = 2 |b|^2 + 2|c|^2
\end{align}
The associated matrix $b^{iso}_{mn}$ that we obtain according to
Eq.~(\ref{eq:relation a_mn b_mn}) is:
\begin{equation}
(b^{iso}_{m,n})= \frac{1}{4}\left(\begin{array}{cccc}
        a+8b+7c&a-c&a-c&a-c\\
        a-c&a-c&a-c&a-c\\
        a-c&a-c&a-8b+7c&a-c\\
        a-c&a-c&a-c&a-c \\
\end{array}\right)
\end{equation}
with a fidelity:
\begin{equation}
 F^{iso}_B=|a|^2+2|b| ^2+|c|^2
\end{equation}
We can now maximize $F^{iso}_B$ for a given $F^{iso}_A$. This is shown in
Fig.~\ref{fig:fidelities}.
For a ``symmetric'' cloner a fidelity of $F^{iso}_A=F_B^{iso}=0,7018$ is
obtained. This is significantly less than the limit threshold of 0,75 that was
obtained by the optimal cloner that clones equally well two Fourier
complementary bases  in
 Ref.~\cite{Cerf02b,Bourennane02}. The difference is
 that  we are now looking at the cloning machine that clones equally
 well (and optimally) two Hadamard complementary bases. In order to
 convince ourselves that they are different cloning machines, it is
 useful to recall that the optimal cloner that was found in
 Ref.~\cite{Cerf02b,Bourennane02}  is symmetric. If we impose this
 symmetry, $a_{m,n}=b_{m,n}$ , at the
 present level, we get the universal or isotropic cloning
machine~\cite{Cerf00b}:

\begin{equation}
(a_{m,n})= \left(\begin{array}{cccc}
        a&b&b&b\\
        b&b&b&b\\
        b&b&b&b\\
        b&b&b&b \\
\end{array}\right)
\end{equation}
The fidelity and disturbances are:
\beqa
F=|a|^2+3|b|^2\\
D_1=D_2=D_3=4|b|^2
\eeqa
The second clone then has a $b_{mn}$ matrix defined as follows
\begin{equation}
(b_{m,n})= \left(\begin{array}{cccc}
\frac{a + 15\,b }{4}&\frac{a - b}{4}& \frac{a - b}{4}&\frac{a - b}{4}\\
  \frac{a - b}{4}&\frac{a - b}{4}&\frac{a - b}{4}& \frac{a - b}{4}\\
  \frac{a - b}{4}&\frac{a - b}{4}&\frac{a - b}{4}&\frac{a - b}{4}\\
  \frac{a - b}{4}&\frac{a - b}{4}&\frac{a - b}{4}& \frac{a - b}{4}\\
\end{array}\right)
\end{equation}
with a fidelity of  $F'=\frac{3}{16}|a-b|^2+\frac{1}{16}|a+15b|^2$.
Due to symmetry and normalization, we have
\begin{align}
a=\frac{\sqrt{10}}{4} \label{eq:a_mn=sqrt 10 over 4} \\
b=\frac{\sqrt{10}}{20} \label{eq:b_mn=sqrt 10 over 20}
\end{align} Note that the fidelity of the optimal
symmetric universal (or isotropic) cloner in N
  dimensions~\cite{Buzek96,Werner98,Cerf00b} is
  $F=\frac{3+N}{2(1+N)}$, which in 4 dimensions yields a fidelity of
  $70\%$. The optimal isotropic quartit cloner
of~\cite{Cerf02b,Bourennane02}, which is
  slightly asymmetric,  is
characterized by a fidelity of $73.33\%$ according to
Ref.~\cite{Bourennane02}, and not $75 \%$ as in
Eq.~(\ref{eq:Fidelity Fourier}).

It is extremely puzzling that two mutually unbiased bases that are
 related by a Hadamard transformation cannot be cloned with the same
 fidelity as bases that are related trough a Fourier transformation.
We found a way to escape this paradox: it consists of a redefinition
of the Bell states.

\subsection{Definition of the Hadamard Bell states}
\label{sec:definition of the hadamar bell states}
It is worth noting that the optimal Fourier cloning state
$\ket{\Psi}^F_{R,A,B,C}$ of Eq.~(\ref{eq:cloneF})  is not invariant  under
arbitrary
permutations. For instance $ x.\sum_{m=0}^{4-1}
\ket{B_{m,0}}_{R,A} \ket{B_{m,0}^*}_{B,C}=x.\sum_{k,i,l=0}^{4-1}\ket{k}_{R}
\ket{k+i}_{A}\ket{l}_{B}\ket{l+i}_{C}$  contains the component
$ x. \ket{0}_{R}\ket{1}_{A}\ket{2}_{B}\ket{3}_{C}$  but not
$ x. \ket{0}_{R}\ket{1}_{A}\ket{3}_{B}\ket{2}_{C}$.
Therefore it is not invariant under permutations of the labels 2 and
3 of the basis states of the computational basis (note that in dimensions
two and three, the corresponding state is invariant under such a relabeling).

Beside, it is easy to check by direct computation that the state
$\ket{\Psi}^F_{R,A,B,C}$  is not invariant in the bases described in
Eqs.~(\ref{base},\ref{base'}). Among others,
the bijective relations described in Eqs.~(\ref{dual1},\ref{dual2}) are not
valid when  the bases are related through a Hadamard transformation.  This
is a novelty
that only appears  in 4 dimensions (or in higher dimensions), because
for lower dimensions one can show
that, up to a convenient redefinition
of the phases of the basis vectors, two mutually unbiased bases are always
the (discrete) Fourier transform of
each other. Essentially this is due to the fact that when the sum of two
(three) phases cancel out, these phases
are unambiguously defined (up to a global phase). This is no longer
true when four phases or more are considered.

If we want to apply the formalism for cloning machines developed
throughout this paper  in order to deal with the
case of two mutually unbiased bases that are not a Fourier
but a Hadamard transformation of
each other, it is
necessary to generalize the definition of a
Bell state. In analogy with the ``Fourier'' cloning state
$\ket{\Psi}^F_{R,A,B,C}$ defined in
Eq.~(\ref{eq:cloneF}), we are now looking for a ``Hadamard'' cloning state
$\ket{\Psi}^H_{R,A,B,C}$ state that is
invariant in the two Hadamard bases, but that
is not necessarily invariant under any permutation of the labels of the
basis states. Moreover, we would be
pleased if a bijective relation similar to those described in
Eqs.~(\ref{dual1},\ref{dual2}) would relate the Bell
states.

To understand how to get the Bell-states that are suited to clone the
Hadamard states, we have to look at the similarities between $F$ (see
Eq.~(\ref{eq:definition F qc})) and
$H$ (see equation~(\ref{adam})). They are both symmetric
matrices, and the elements of their
columns (rows) form a group with the product operator. Indeed we have
\begin{equation}
2F_{i,j} \,\, 2F_{i,k}=2F_{i,j+k}
\label{eq:fourier row group}
\end{equation}
where the sum is (as before) modulo 4.
We want to express the group structure of $H$ in a similar
way. Therefore we introduce a new `Hadamard sum' operator:
\begin{align}
&i\oplus j = (i + j) \mod 4\\ \intertext{ except when both $i$ and $j$
are equal to 3 or 1 }
&1\oplus 1=3\oplus3=0 & & 1\oplus 3=3\oplus1=2
\end{align}
With this definition we can write as in~(\ref{eq:fourier
  row group}):
\begin{equation}
2H_{i,j} \,\, 2H_{i,k}=2H_{i,j\oplus k}
\label{eq:hadamard row group}
\end{equation}

We have also shown in the section \ref{sixa} that if  we permute
cyclically the labels of the computational basis, the states of the
Fourier basis were mapped onto  themselves and vice versa. This leads
to the definition the Fourier Bell states:
\begin{equation}
B_{m,n}^F=\sum_{k=0}^3 F_{k,n} C^k \otimes C^{k+m}
\ket{00}=\sum_{k=0}^3 F_{k,n} \ket{k}\ket{k+ m}
\label{eq:Bell states fourier}
\end{equation}
with C the operator that performs the cyclic permutation.
The Hadamard states do not have this cyclic symmetry. But there are
other permutation symmetries  in this case. We define three permutations in the
computational basis (the same permutations in the primed basis are
noted with a prime): $P_1$
switches $0 \leftrightarrow 1$ and simultaneously $2 \leftrightarrow 3$, $P_2$
switches $0 \leftrightarrow 2$  and simultaneously  $1 \leftrightarrow
3$ and $P_3$ switches $0 \leftrightarrow 3$  and simultaneously  $1
\leftrightarrow 2$. Together with the identity (we will call $P_0\equiv
I$) these permutations form a commutative group. It is obvious that
$P_i$ ($P'_i$) maps the basis states of the primed (non-primed) basis
onto themselves (up to a global phase). Completely analogous
with~(\ref{eq:Bell states fourier}) we define the Hadamard Bell
states:
\begin{equation}
B_{m,n}^H=\sum_{k=0}^3 H_{k,n} P_k \otimes P_{k\oplus m}
\ket{00}=\sum_{k=0}^3 H_{k,n} \ket{k} \ket{k\oplus m}
\label{eq:Bell states Hadamar}
\end{equation}

The 16 Hadamard Bell states are contained in the following list. They
form a maximally entangled orthonormal basis, and have still all the
interesting properties that characterize Fourier Bell states (complementarity
between Eve's and Bob's reduced density operators, interpretation of
the cloning state in terms of error operators, \ldots) as we shall show
soon. The parities
under the permutations
$P_1$,
$P_3$, $P_1'$, and $P_3'$ are given in parentheses for each state.
\begin{align} \label{adambell}
\ket{B^H_{0,0}}=\frac{1}{2}(\ket{00}+\ket{11}+\ket{22}+\ket{33})(+++'+'),\ket{B^
H_{0,1}}=\frac{1}{2}(\ket{00}+\ket{11}-\ket{22}-\ket{33})(+-+'+')\nonumber\\
\ket{B^H_{0,2}}=\frac{1}{2}(\ket{00}-\ket{11}+\ket{22}-\ket{33})(--+'+'),\ket{B^
H_
{0,3}}=\frac{1}{2}(\ket{00}-\ket{11}-\ket{22}+\ket{33})(-++'+')\nonumber\\
\ket{B^H_{1,0}}=\frac{1}{2}(\ket{01}+\ket{10}+\ket{23}+\ket{32})(+++'-'),\ket{B^
H_{1,1}}=\frac{1}{2}(\ket{01}+\ket{10}-\ket{23}-\ket{32})(+-+'-')\nonumber\\
\ket{B^H_{1,2}}=\frac{1}{2}(\ket{01}-\ket{10}+\ket{23}-\ket{32})(--+'-'),\ket{B^
H_
{1,3}}=\frac{1}{2}(\ket{01}-\ket{10}-\ket{23}+\ket{32})(-++'-')\nonumber\\
\ket{B^H_{2,0}}=\frac{1}{2}(\ket{02}+\ket{20}+\ket{13}+\ket{31})(++-'-'),\ket{B^
H_{2,1}}=\frac{1}{2}(\ket{02}-\ket{20}+\ket{13}-\ket{31})(+--'-')\nonumber\\
\ket{B^H_{2,2}}=\frac{1}{2}(\ket{02}+\ket{20}-\ket{13}-\ket{31})(---'-'),\ket{B^
H_
{2,3}}=\frac{1}{2}(\ket{02}-\ket{20}-\ket{13}+\ket{31})(-+-'-')\nonumber\\
\ket{B^H_{3,0}}=\frac{1}{2}(\ket{03}+\ket{30}+\ket{12}+\ket{21})(++-'+'),\ket{B^
H_{3,1}}=\frac{1}{2}(\ket{03}-\ket{30}+\ket{12}-\ket{21})(+--'+')\nonumber\\
\ket{B^H_{3,2}}=\frac{1}{2}(\ket{03}-\ket{30}-\ket{12}+\ket{21})(---'+'),\ket{B^
H_
{3,3}}=\frac{1}{2}(\ket{03}+\ket{30}-\ket{12}-\ket{21})(-+-'+')
\end{align}

We can now easily show that we have a bijective relation similar
to~(\ref{dual1},\ref{dual2}):
\begin{align}
\ket{B^H_{i,j}}&=\sum_{l=0}^{3}H_{l,j}\ket{l,i\oplus l}\\
&= \sum_{k,l,m,p=0}^{3}H_{l,j}H_{l,m}H_{i\oplus
l,p}\ket{m',p'} =\sum_{k,l,m,p=0}^{3}2H_{l,j}H_{l,m}H_{i,p}H_{l,p}\ket{m',p'} \\
&=  \sum_{l,m,p=0}^{3}H^{-1}_{p,l}H_{l,j\oplus
m}H_{i,p}\ket{m',p'}= \sum_{m,p=0}^{3}\delta_{p,j\oplus
  m}H_{i,p}\ket{m',p'}\\
&= \sum_{m=0}^{3}H_{i,j\oplus m}\ket{m',(j\oplus
  m)'}= \sum_{m=0}^{3}2H_{i,j}H_{i, m}\ket{m',(j\oplus
  m)'}\\
\ket{B^H_{i,j}}&=2H_{i,j}\ket{B'^H_{j,i}}
\end{align}

Similarly, we can evaluate the in-product between $\ket{B^H_{i,j}}_{R,A}
\ket{B^H_{i,j}}_{B,C}$ and $\ket{B^H_{m,n}}_{R,B}
\ket{B^H_{m,n}}_{A,C}$ as follows:
\begin{align}
\braket{B^H_{m,n R,B} &.B^H_{m,n A,C}}{B^H_{i,j R,A}.B^H_{i,j B,C}}\\
&=\sum_{o,p,l,k=0}^{3}H_{o,n}H_{p,n}H_{l,j}H_{k,j}\braket{o,o\oplus
m,p,p\oplus m} {l,k,l\oplus i,k \oplus i }\\
&=\sum_{o,p,l,k=0}^{3}H_{o,n}H_{p,n}H_{l,j}H_{k,j}\delta_{o,l}
\delta_{o \oplus m,k}\delta_{p,l\oplus i }\delta_{p \oplus
m,k\oplus i}\\
&=\sum_{l=0}^{3}H_{l\oplus i,n}H_{l,n}H_{l\oplus m,j}H_{l,j}\\
&=\sum_{l=0}^{4-1}4H_{i,n} H_{l,n}H_{l,n}H_{
m,j}H_{l,j}H_{l,j}\\
&=H_{i,n} .H_{ m,j}
\end{align}

Note that this defines a unitary transformation, since the
states $\ket{B^H_{i,j}}_{R,A}
\ket{B^H_{i,j}}_{B,C}$ and $\ket{B^H_{m,n}}_{R,B}
\ket{B^H_{m,n}}_{A,C}$ are bases of the space that is invariant or
symmetric (eigenvalues +) under the permutations
$P_i$ and $P'_j$ (i,j=0,1,2,3). Now that we know the in-products between these
states, it is very easy to derive the
following relation:
\beq \label{eq:jointstate2}\ket{\Psi}^H_{R,A,B,C}=\sum_{m,n=0}^{3} a_{m,n}
\; \ket{B^H_{m,n}}_{R,A}
\ket{B^H_{m,n}}_{B,C} = \sum_{m,n=0}^{3} b_{m,n} \;
\ket{B^H_{m,n}}_{R,B} \ket{B^H_{m,n}}_{A,C}
\eeq
where $a_{m,n}$ and $b_{m,n}$ are two (complex) amplitude functions
that are dual under a Hadamard transform (which generalizes the dual
relation Eq.~(\ref{eq:relation a_mn b_mn})):
\begin{align}  \label{HT}
b_{m,n} &= \sum_{x,y=0}^{3} H_{m,y}H_{n,x}a_{x,y} \\
a_{m,n} &= \sum_{x,y=0}^{3} H_{m,y}H_{n,x}b_{x,y}
\end{align}
Also the error operators, which for the Fourier cloner are defined as:
\begin{align}
U_{m,n}^F = 2 \sum_{k=0}^3 F_{k,n} \ket{k+m} \bra{k}\\
\intertext{are redefined as:}
U_{m,n}^H = 2 \sum_{k=0}^3 H_{k,n} \ket{k\oplus m} \bra{k}
\end{align}

With these new definitions, the analysis of the cloner is analogous to
the treatment of the cloner presented in Sec.~\ref{sec:Applying cov form to fourier} 
and in~\cite{Cerf02b,Bourennane02,Cerf00b}.
The reduced density operator the two clones are:
\begin{align}
\rho_A&=\sum_{m,n} a_{m,n} \proj{\psi_{m,n}^H}\\
\rho_B&=\sum_{m,n} b_{m,n} \proj{\psi_{m,n}^H}
\end{align}
with
\begin{equation}
\ket{\psi_{m,n}^H}=U^H_{m,n}\ket{\psi}
\end{equation}
The fidelities and disturbances are given by Eq.~(\ref{eq:F_A in computational basis}) and Eq.~(\ref{eq:disturbances in the computational basis}).  Therefore,
the optimal cloner will have the same amplitudes as
in Sec.~\ref{sec:Applying cov form to fourier}
and~\cite{Cerf02b,Bourennane02,Cerf00b}:
where \begin{equation}
(a_{m,n})=(b_{m,n})=(a'_{m,n}) =\left(\begin{array}{cccc}
x&y&y &y\\
y&z& z&z\\
 y&z& z&z\\
y&z& z&z\\
\end{array}\right)
\end{equation}

With these amplitudes, the state of Eq.~(\ref{eq:jointstate2}) is the
optimal state that clones the computational and the Hadamard bases equally
well.

The reduced state
obtained after tracing over Eve's degrees of freedom  exhibits no
correlations when it
is measured by Alice and Bob in
non-correlated bases, so that it is impossible for them to establish the
difference between this state and an
unbiased noise.

The corresponding maximal admissible error
rate (when attacks based on state-independent cloners are considered)
 can be
shown as before to be equal to $E_A=1-F_A =\frac{1}{2}(1-\frac{1}{\sqrt{4}})=
25.00\%$ (see
 Ref.~\cite{Cerf02b,Bourennane02}). According to Csisz\'{a}r and K\"{o}rner
theorem\cite{Csiszar78}, the quantum
cryptographic protocol above ceases to generate secret key bits precisely
at this point, where Eve's information
matches Bob's information.

\section{Conclusion and comments}
\label{sec:conclusion}
In this paper we investigated state dependent cloners that are suited
to clone four-level systems (quartits). We have adapted the cloning
formalism of N.Cerf~\cite{Cerf00,Cerf00b} in such a way that it is
covariant under
certain unitary transformations.
We have used this protocol to clone the states of two quartit bases that
are related
through a double Hadamard transform. Our results show that the protocol
is not suited to clone such states, since the cloning state uses
`Fourier' Bell states which have different symmetries than the Hadamard
bases that we want to clone. Therefore, we redefined the Bell states in
such a way
that they have the same symmetries as the states that we want to
clone. The result is a cloner  that  clones the Hadamard
bases equally well as the 'Fourier' cloner clones Fourier bases. Note
that the approach is suited for an attack on all protocols that use
Hadamard bases to encode the information as in Ref.~\cite{Bechmann99b,Brainis03}.

 The question remains open whether better cloning machines exist. The
safety of quantum
cryptographic protocols depends crucially on the answer. It is out of the
scope of the present paper to provide a
definitive answer to this question, but it is worth mentioning that in the
qubit ($N$=2) and qutrit ($N$=3) cases, the
properties (optimal fidelity, upper bound on the error rate and so on) of
the cloners derived in the literature
following Cerf's approach~\cite{Cerf02b,Bourennane02,Cerf02c,Durt03} are
equivalent to those obtained following more
general approaches~\cite{Fuchs97,Bruss00,Bechmann99,Kaszlikowski02}.

Many constraints have to be fulfilled when Eve
replaces the maximally entangled state
shared by Alice and Bob by a clone: the clone must mimic the unbiased
noise that would be observed by Alice and Bob in the
absence of a spy. Therefore it must be independent on the basis of
detection and on which state is detected among
such a basis. Moreover it must mimic the correlations between different
bases that would be observed in the
presence of unbiased noise.

It is our belief that the present, constructive, approach, based on the
intrinsic symmetries of the protocols
under study, provides the most dangerous attack, taking account of all the
constraints of the problem.

\begin{acknowledgments}

T.D. is a
Postdoctoral Fellow and B.N. research assistant of the Fonds voor
Wetenschappelijke Onderzoek-Vlaanderen.
This research was supported by the Belgian Office for Scientific, Technical
and Cultural Affairs in the framework of the Inter-University Attraction Pole
Program of the Belgian
governement under grant V-18., the Fund for Scientific Research - Flanders
(FWO-V), the
Concerted Research
Action ``Photonics in Computing'' and  the research council (OZR) of the
VUB.
After the completion of this work, we were informed (N. Cerf private
communication) that N. Cerf and his coworkers obtained independently
certain results similar to ours. Thanks to N.
Cerf, S. Iblisdir, L-P Lamoureux, S. Massar and P. Navez for interesting
discussions.
\end{acknowledgments}

\end{document}